\documentclass[letterpaper, onecolumn]{article} % DO NOT CHANGE THIS
\usepackage{aaai21}  % DO NOT CHANGE THIS
\usepackage{times}  % DO NOT CHANGE THIS
\usepackage{amsmath}
\usepackage{amssymb}

\usepackage{helvet} % DO NOT CHANGE THIS
\usepackage{courier}  % DO NOT CHANGE THIS
\usepackage[hyphens]{url}  % DO NOT CHANGE THIS
\usepackage{graphicx} % DO NOT CHANGE THIS
\usepackage{natbib}  % DO NOT CHANGE THIS AND DO NOT ADD ANY OPTIONS TO IT
\usepackage{caption} % DO NOT CHANGE THIS AND DO NOT ADD ANY OPTIONS TO IT
\usepackage{url}                                %Required
\usepackage{array,graphicx,adjustbox}  %Required
\usepackage{booktabs}                           % For formal tables
\usepackage{color}
\usepackage{subfigure}
\usepackage{paralist}
\usepackage{enumitem}
\usepackage{tikz}

\usepackage{xcolor}
\usepackage{makecell} 
\usepackage{multicol}
\usepackage{multirow}

\nocopyright
\graphicspath{{./figs/}}

\definecolor{myblue}{rgb}{0.333, 0.447, 0.878}
\definecolor{myred}{rgb}{0.816, 0.286, 0.247}
\definecolor{myrederr}{rgb}{0.894, 0.703  , 0.699}
\definecolor{grey}{rgb}{0.2, 0.2, 0.2}

\newcommand{\ms}{in} 

\def\PosbarErr[#1]#2{~#1 {
\begin{minipage}{2.4\ms}
\begin{flushleft}
{\color{white}\rule{0.8\ms}{0pt}}%
{\color{myred}\rule{#1\ms}{6pt}}%
{\color{grey}\rule[2.5pt]{#2\ms}{1pt}}%
\end{flushleft}
\end{minipage}%}
} }

\def\NegbarErr[#1]#2{~#1 {
\begin{minipage}{2.4\ms}
\begin{flushright}
{\color{grey}\rule[2.5pt]{#2\ms}{1pt}}%
{\color{myblue}\rule{-#1\ms}{6pt}}%
{\color{white}\rule{1.6\ms}{0pt}}%
\end{flushright}
\end{minipage}
} }

\title{Mapping urban socioeconomic inequalities in developing countries through Facebook advertising data}
\author {
    Serena Giurgola,\textsuperscript{\rm 1,\thanks{Authors contributed equally}}
    Simone Piaggesi, \textsuperscript{\rm 1, 2, \footnotemark[1]}
    M\'arton Karsai, \textsuperscript{\rm 3, 1}\\
    Yelena Mejova, \textsuperscript{\rm 1} 
    André Panisson, \textsuperscript{\rm{1}}
    Michele Tizzoni \textsuperscript{\rm{1}}\\
}
\affiliations {
    \textsuperscript{\rm 1} ISI Foundation, Turin, Italy \\
    \textsuperscript{\rm 2} Università degli Studi di Bologna, Bologna, Italy\\
    \textsuperscript{\rm 3} Central European University, Vienna, Austria\\

}

\begin{document}
\maketitle

\begin{abstract}
  Ending poverty in all its forms everywhere is the number one Sustainable Development Goal of the UN 2030 Agenda. 
  To monitor the progress towards such an ambitious target, reliable, up-to-date and fine-grained measurements of socioeconomic indicators are necessary. 
  When it comes to socioeconomic development, novel digital traces can provide a complementary data source to overcome the limits of traditional data collection methods, which are often not regularly updated and lack adequate spatial resolution. 
  In this study, we collect publicly available and anonymous advertising audience estimates from Facebook to predict socioeconomic conditions of urban residents, at a fine spatial granularity, in four large urban areas: Atlanta (USA), Bogotá (Colombia), Santiago (Chile), and Casablanca (Morocco). 
  We find that behavioral attributes inferred from the Facebook marketing platform can accurately map the socioeconomic status of residential areas within cities, and that predictive performance is comparable in both high and low-resource settings. 
  We also show that training a model on attributes of adult Facebook users, aged more than 25, leads to a more accurate mapping of socioeconomic conditions in all cities. 
  Our work provides additional evidence of the value of social advertising media data to measure human development. 
  
\end{abstract}

\section{Introduction}

Reduction of poverty is the number one goal of the United Nations, as defined in their Sustainable Development Goals (SDG)\footnote{\url{https://sdgs.un.org/goals}}.
However, in order to address this age-old condition, copious amounts of data need to be collected, and often it is in the places most at risk that it is most difficult to perform surveys. 
In 2020, the World Bank has admitted that surveying and face-to-face interviewing have been hindered by the COVID-19 epidemic and the resulting distancing measures\footnote{\url{https://www.worldbank.org/en/topic/measuringpoverty}}. 
Technology-assisted surveys, including phone-based interviews, are becoming invaluable tools. However, many countries lack the resources to run such data collection exercises, and monitor the socioeconomic status (SES) indicators only every several years, some only every decade.

Cities are remaining at the center of these developments, as urban populations grow rapidly, and are projected to do so in the near future \cite{dfid2021meeting,jha2020how}. 
Urbanization contributes to global economic growth, provides opportunities for millions of people, attracts investors and entrepreneurs, and offers much needed services \cite{baker2008urban}.
However, it also suffers from maladies spanning overcrowding and inadequate housing, lack of social networks, stark inequality, crime, and violence. 
In some cases, undocumented residents forego the benefits of urbanization and miss out on government assistance, such as during the COVID-19 pandemic, when the joblessness disproportionally affected non-white and female workforce \cite{urban2020immigrant}. 
In such dynamic circumstances, the well-being of urban residents often hinges on the inclusivity, infrastructure, and other measures taken by their urban governance systems \cite{world2020urban}.

To keep track of the urban SES indicators, recent research has turned to alternative data sources, including the daytime \cite{jean2016combining,watmough2019socioecologically,engstrom2017poverty} and nighttime satellite imagery \cite{chen2011using,mellander2015night}, mobile phone Call Detail Records (CDRs) \cite{fernando2018predicting}, and even crowd-annotated information from OpenStreetMap \cite{tingzon2019mapping}.
However, even these data sources often suffer from being out of date and not easily available to the research community.
In this paper, we use a resource which has been gaining attention in the demographic domain: the advertising platform of the largest social network, Facebook, accessible via the Facebook Graph API\footnote{\url{https://developers.facebook.com/docs/marketing-apis/}}. 
In particular, Facebook provides a way to gauge the reach of any advertising campaign by providing the number of daily or monthly active users (MAU) that certain constrains would reach. 
For instance, it is possible to target Facebook users by their gender, age, sets of interests, location (down to a circle of 1 km radius), and many other attributes\footnote{\url{https://developers.facebook.com/docs/marketing-api/reference/ad-campaign-delivery-estimate/}}. 
These estimates provide a ``view'' of the billions of Facebook users without jeopardizing the privacy of any individual user, and they can be obtained automatically, without running the actual ads. 
Recently, Facebook Ads audience estimates have been used to track the prevalence of obesity and diabetes \cite{araujo2017using}, crime rates \cite{fatehkia2019correlated}, cultural assimilation \cite{stewart2019rock} and mass migration \cite{palotti2020monitoring}.
Socioeconomic development across the Indian states has been tracked using the gender disparities in the access to Facebook \cite{mejova2018measuring}, and in combination with satellite imagery it has been applied to Philippines and India for tracking Demographic and Health Survey Wealth Index \cite{fatehkia2020relative}. 
However, it is still unclear how applicable this information is at a fine granularity, and whether adoption rates of Facebook in the population would render Facebook too sparse for developing urban regions. 

In this paper, we seek to fill this gap by answering the following research questions:
\begin{itemize}
    \item {\bf RQ1}: can Facebook advertising audience estimates provide insights into the socioeconomic conditions of populations at a high spatial granularity, namely at the level of urban subdivisions? 
    
    \item {\bf RQ2}: is the performance of a predictive model at such spatial resolution comparable across cities in high and low-income economies, characterized by different Facebook penetration levels? 
    
    \item {\bf RQ3}: what Facebook users' attributes are most predictive of socioeconomic status within urban areas?
\end{itemize}
To this aim, we use Facebook audience estimates to predict SES of the neighbourhoods of large urban areas: Atlanta (USA), Santiago (Chile), Bogotá (Colombia), and Casablanca (Morocco). 
We choose cities in both developed an developing countries, to gauge the effectiveness of our approach in different settings characterized by middle-low to high income economies.
Through the Facebook marketing API, we measure the spatial distribution of monthly active users matching a wide range of targeting options, including demographics, behaviors and interests, within each city.
We consider audience estimates disaggregated into two age groups: all Facebook users and adults, or older than 25, only.  
We then use such estimates to predict SES, defined as annual income or poverty rate, of neighbourhoods' residents.

Our results demonstrate that Facebook audience estimates can provide insights into the spatial patterns of wealth and poverty in urban areas, with good and comparable predictive performance between developed and developing countries. 
Furthermore, we show that poverty predictions can significantly improve when a model is trained on audience estimates of adult users only, especially in resource-poor settings.

% %%% %%% %%% %%% %%% %%% %%% %%% %%% %%% %%% %%% %%% %%% %%% %%% %%% %%% %%%

\section{Related work}

The Data Revolution has made available not only existing records in a digitized form, but it has spurred an interest in alternative sources of data \cite{alburez-gutierrez_zagheni_aref_gil-clavel_grow_negraia_2019, weber2021non}. Below, we summarize the latest efforts in using alternative digital signals, and social media in particular, in order to track SES.  

The socioeconomic status of individuals is a complex character, which depends not only on one's economic capacities but also on the social and cultural position of the ego in the larger society. Quantifying such a convoluted character of a person is a very difficult if not impossible task~\cite{oakes2003measurement,baumeister2007encyclopedia,vyas2006constructing}. 
For this reason, data-driven studies usually approximate socioeconomic status by some easily observable variables, which sensitively reflect economic inequalities between people. 
Such indicators can be the income~\cite{abitbol2020interpretable} or the occupation~\cite{abitbol2018location} of an individual, or poverty level of one’s residential neighborhood~\cite{steele2017mapping}, just to mention a few examples. 
Meanwhile, for a meaningful analysis these indicators are needed to be available for larger populations as they commonly serve as ground truth data for supervised inference methods. For behavioural based inference individual indicators correlated with one's economic capacity are used, like bank~\cite{leo2016socioeconomic} and purchase records~\cite{leo2016correlations}. 
At the same time, location-based inference requires high-resolution income and demography maps typically recorded during census in developed countries~\cite{insee}, or low-resolution poverty maps from under-developed countries are used for these purposes~\cite{wbpoverty}.

A variety of digital sources have been used to track development and SES, including satellite imagery \cite{elvidge2009global, piaggesi2019predicting}, mobile call log data \cite{blumenstock2015predicting}, and transport-related apps \cite{tan2019neighborhood}.
Social media in particular has been used to provide deeper understanding of various population well-being indices. For instance, Resce \& Maynard \cite{resce2018matters} use Twitter to track the constituent issues comprising the Better Life Index (BLI) including income, employment, civic engagement, and health. Google Trends, a service providing an aggregated view of common Google search queries, has been used to track infectious and non-communicable diseases \cite{nuti2014use}, as well as unemployment and consumer confidence \cite{choi2012predicting}. Moreover, recent inference methods provide high-resolution estimates of poverty maps even in under-developed countries using combined data sources and validation on spatially aggregated levels~\cite{lee2020high}.

% Focus on benefits of social media
Compared to traditional data sources, social media offers several notable benefits. Due to the data being available in real time via Application Programming Interfaces (APIs) by the platforms, it can be used to provide rough estimates of ongoing phenomena, or help in ``nowcasting'' \cite{di2018big}. 
As official data often lags by as much as weeks or months, nowcasting using social media provides daily and even hourly information in volatile situations including disasters and ongoing events. Coverage is another benefit of social media -- especially either of large platforms that have been widely adopted, or of smaller, more local platforms. 
For example, searches on Baidu have been used to estimate economic activity \cite{dong2017measuring} and restaurant reviews to estimate socioeconomic attributes of urban neighborhoods \cite{dong2019predicting} in China. 
However, the full datasets are usually not available for research, and the APIs provide a small glimpse into the vast user bases of major social media websites. Their advertising services, then, are an alternative route to learning about the users of large websites in a privacy-preserving fashion.

% Socio-economic indicators thru Facebook Ads
Facebook (and most other large websites) provides advertising services on its platform which allow potential advertisers to ascertain the size of the potential audience their campaign could reach. 
Along with the basic demographic, location, and technology use, it provides the advertisers to explore their audiences by a variety of interests and behaviors. Recently, demographers, sociologists, and other researchers have been using this information as a kind of ``digital census''.  
Using Facebook Ads, a variety of demographic and economic indicators have been studied, such as the prevalence of obesity and diabetes \cite{araujo2017using}, crime rates \cite{fatehkia2019correlated}, cultural assimilation \cite{stewart2019rock} and mass migration \cite{palotti2020monitoring}. 
However, the usefulness of such data may vary in different locales, especially compared to alternative sources of information. For instance, \cite{fatehkia2020relative} show that models trained on Facebook Ads data can predict the Demographic and Health Survey Wealth Index in Philippines about as well as those trained on satellite data. However, for India, satellite data performs better, possibly due to the lower penetration of Facebook.
Especially useful may be the signals about the kind of technology that is available to the populations, such as mobile phones and network access \cite{fatehkia2020mapping}. 
Recent work has also shown that ads audience data provided by Facebook suffers from inconsistency over time and poor coverage in sparsely-populated areas \cite{rama2020facebook}. Still, the same work has shown that it is possible to overcome some of these challenges, and to capture multiple dimensions of inequality between rural and urban municipalities in Italy \cite{rama2020facebook}.
In this study, we examine the usefulness of Facebook Ads audience estimates both in developed and developing urban settings, and discuss the challenges and benefits it brings with regard to mapping the fine-grained SES levels.

\section{Data collection and methods}

In this section, we describe the main data sources analyzed in our study and the methods used to predict the urban SES indicators. To aid in the reproducibility of the study, we will release the data and code along with the final version of the paper. 

\subsection{Urban socioeconomic indicators}

%%%FIGURE 1%%%%%
\begin{figure}[t]
  \centering
  \includegraphics[width=\columnwidth]{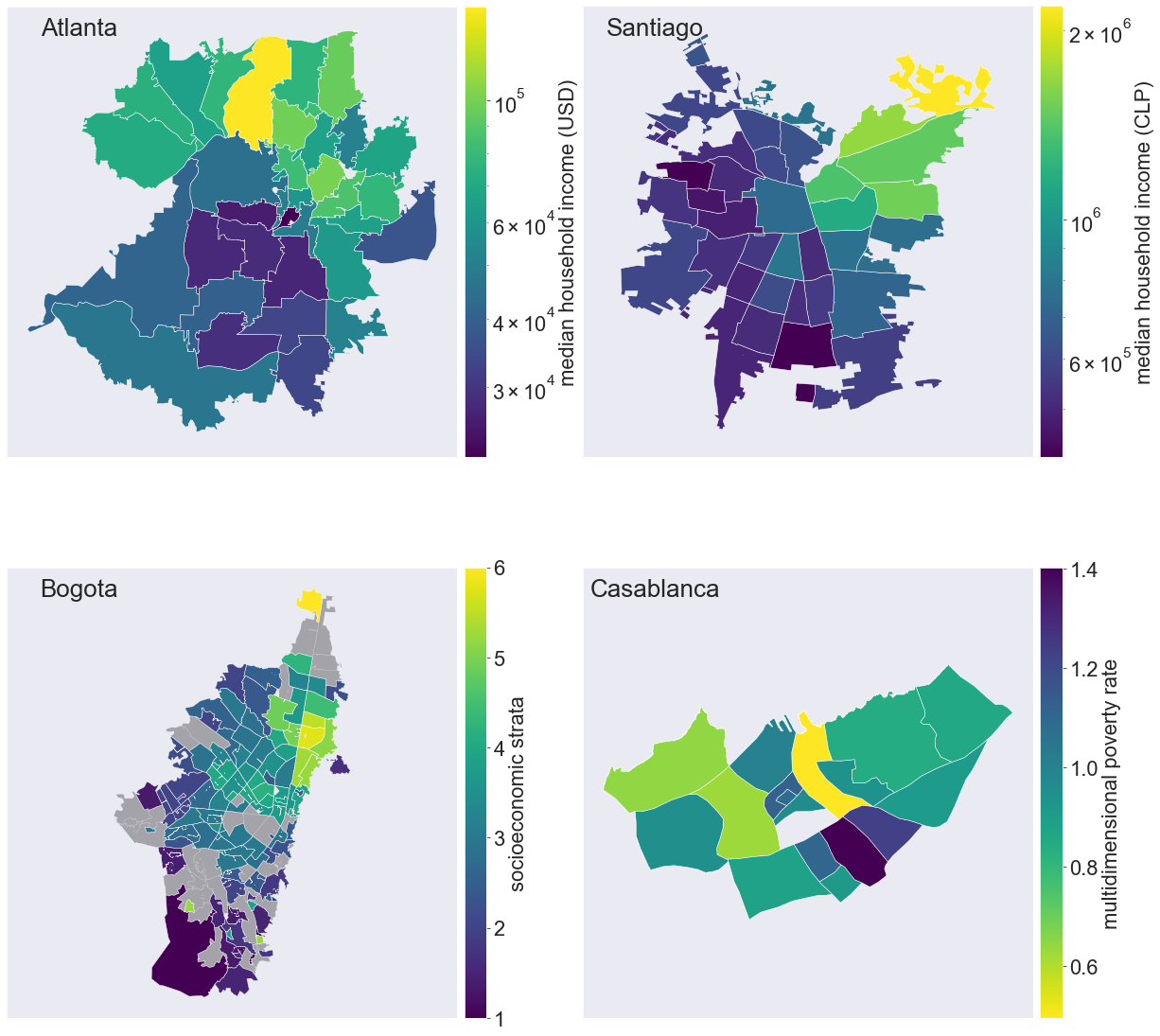}
  \caption{ \label{fig:maps} Spatial distributions of socioeconomic indicators. Choropleth maps display the values of socioeconomic indicators in the cities of Atlanta, Santiago, Bogotá and Casablanca. From top to bottom: median household income in Atlanta, USA, by zip code; median household income in Santiago, Chile, by comuna; socioeconomic strata in Bogotá, Colombia, by UPZ (missing data are colored in gray); multidimensional poverty rate, in Casablanca, Morocco, by arrondissement.}
\end{figure}

\begin{table*}
\begin{center}
  \caption{Socioeconomic indicators used in the study and their corresponding data sources. The rightmost column indicates the number of data points, corresponding to the number of neighborhoods in each city.}
  \label{tab:indicators}
  \begin{tabular}{lcccl}
    \toprule
    City & Indicator & Unit &  Year & $N$\\
    \midrule
    Atlanta, USA & \makecell{Median household \\ income} & US Dollars (USD) & 2019 & 40\\
    
    Bogotá, Colombia & Socioeconomic strata & Levels 1 (low) - 6 (high)  & 2017 & 110\\
    
    Santiago, Chile & Median household income  & Chilean Pesos (CLP)  & 2012 & 36\\
    
    Casablanca, Morocco & Multidimensional poverty rate & Population fraction  & 2014 & 17\\
  
  \bottomrule
\end{tabular}
\end{center}
\end{table*}

We began by collecting socioeconomic data in the cities of Atlanta, Santiago, Bogotá, and Casablanca from publicly available official sources. 
Tab.~\ref{tab:indicators} summarizes the main characteristics of the data sets under study. 
The four data sets map the socioeconomic status of the neighbourhoods in each city based on different indicators and at different spatial resolutions. 
In our study, these represent our main target that we aim to predict using advertising audience estimates as described below.

In the city of Atlanta, we considered the median household income in each of the 40 zip codes, as reported by the American Community Survey in 2019 \cite{CensusReporter}.
In Santiago, Chile, we collected the 2012 median household income, in Chilean Pesos, reported by the Chilean Ministry of Transport and Communication \cite{EOD}. 
We considered the urban part of the Santiago Metropolitan Area, that is composed of 36 municipalities, named \emph{comunas}. 
Socioeconomic data for Bogotá are available from the \emph{Secretar\'ia Distritial de Planeaci\'on} and map the socioeconomic status of 110 neighborhoods (Unidades de Planeamiento Zonal or UPZ) on a discrete scale ranging from 1 (low income) to 6 (high income).  
Finally, in the 17 neighbourhoods (\emph{arrondissements}) of Casablanca, we considered the multidimensional poverty rate reported by the official census in 2014. 
The multidimensional poverty rate measures the population fraction living in poverty, according to the definition of the the Higher Planning Commission of Morocco \cite{HCP} which takes into account 15 different dimensions, ranging from income to education, health and access to essential services. 
The spatial distribution of the socioeconomic indicators in the neighbourhoods of Atlanta, Bogotá, Santiago and Casablanca is shown in Fig.~\ref{fig:maps}. 
In all cities, we can observe distinct geographic patterns with strong socioeconomic inequalities across districts.
In Atlanta, the 2019 median household income ranged from 30,000 USD or less in Downtown and southern areas of the city, to 100,000 USD or more in the wealthiest residential areas of the North. 
Segregation patterns in Santiago follow a East-West divide, where the wealthiest neighborhoods are clustered in the Northeastern part of the city. 
A median household income above 1 million Chilean pesos is observed in only 5 out of 36 \emph{comunas}.
In Bogotá, higher socioeconomic strata of the population are concentrated in the neighbourhood of Usaquén, in the North-East of the city.
In Casablanca, the most deprived \emph{arrondissements}, with a poverty rate above 1\%, are found in the South-East suburbs of the city.

\subsection{Facebook advertising data}

We collected Facebook Advertising audience estimates through the Facebook Marketing API using the Facebook business Python package\footnote{\url{https://github.com/facebook/facebook-python-business-sdk}}. 
Each query targets a geographic area of interest, either a city neighborhood or a circle with a fixed radius of 1 km. 
More details about the spatial definition of our queries are provided in subsection \ref{sec:spatial}.
In each query, we request the count of Facebook users ``who live there'' (technically, by setting the \texttt{location\_type} parameter to \texttt{home}), and who match a specific targeting option as described below.
We constrain each query to select only Facebook users, although it is possible to query for users using other Facebook owned services, like Instagram. 
Given our interest in different countries, in this way we aim at an easier comparison and interpretation of results.  
Among the various advertising campaign types, we choose the ``reach'' option, which targets the ``maximum number of people''. 
Finally, in the reply to our query, we save the number of Monthly Active Users (MAU), as done in previous studies, because it provides a more stable estimate with respect to the Daily Active Users \cite{rama2020facebook}. 

Since each query does not usually return exactly the same response from the API \cite{rama2020facebook}, especially in less populated areas, we average all audience estimates over 3 identical queries, performed at different times, between January and September 2020. 
The Facebook marketing platform does not return estimates of MAU below the value of 1000, and if a query targets a smaller number of users, then the API will return a value of 1000 - which is thus indistinguishable from 0. 
In our study, we replaced all query results equal to 1000 with zeros and when a specific combination of target and location returned 1000 users for all the 3 queries, we did not include it among the features of the predictive models.    

\subsection{Targeting options}

\begin{table*}
\begin{center}
  \caption{List of Facebook attributes considered as predictive features of SES. For each city, we show with a check mark if a feature was included or not as input into the regression model without age limits, before variable selection. Features were included only if the corresponding users estimate did not hit the 1000 users threshold in all districts of a city.}
  \label{tab:features}
  \begin{tabular}{lrcccc}
    \toprule
    Category &  Feature & Atlanta & Bogotá & Santiago & Casablanca \\
    \midrule
    
    \multirow{2}{*}{Gender} & \texttt{Males} & \checkmark  & \checkmark  & \checkmark  & \checkmark \\
                            &  \texttt{Females}  & \checkmark  & \checkmark  & \checkmark  & \checkmark \\
    \midrule
    \multirow{4}{*}{Marital status}           & \texttt{Single}  & \checkmark & \checkmark & \checkmark & \checkmark\\ 
                                                &  \texttt{Engaged}  & \checkmark & \checkmark & \checkmark & \checkmark\\
                                                 & \texttt{Married}  & \checkmark & \checkmark & \checkmark & \checkmark\\
                                                 & \texttt{Civil Union}  &  & \checkmark & \\
    \midrule
    \multirow{2}{*}{Education}           & \texttt{High school grad}  & \checkmark & \checkmark & \checkmark & \checkmark\\ 
                                                &  \texttt{College grad}  & \checkmark & \checkmark & \checkmark & \checkmark\\
    \midrule
    \multirow{7}{*}{Travel}           & \texttt{Away from hometown}  & \checkmark & \checkmark & \checkmark & \checkmark \\ 
                                                &  \texttt{Away from family}  & \checkmark & \checkmark & \checkmark & \checkmark \\
                                                &  \texttt{Frequent international travelers}  & \checkmark & \checkmark  & \checkmark & \checkmark\\
                                                &  \texttt{Frequent travelers}  & \checkmark & \checkmark & \checkmark & \checkmark\\
                                                &  \texttt{Return from travels 1 week ago}  & \checkmark & \checkmark & \checkmark & \checkmark\\
                                                &  \texttt{Expats}  & \checkmark & \checkmark & \checkmark & \checkmark\\
    \midrule
    \multirow{8}{*}{Interests}           & \texttt{Gambling}  & \checkmark & \checkmark & \checkmark & \checkmark\\ 
                                                &  \texttt{Casino}  & \checkmark & \checkmark & \checkmark & \checkmark\\
                                            & \texttt{Cooking}  & \checkmark & \checkmark & \checkmark & \checkmark\\ 
                                                &  \texttt{Restaurants}  & \checkmark & \checkmark & \checkmark & \checkmark\\
                                                &  \texttt{Fast food}  & \checkmark & \checkmark & \checkmark & \checkmark\\
                                                &  \texttt{Fitness and wellness}  & \checkmark & \checkmark & \checkmark & \checkmark\\
                                                &  \texttt{LGBT community}  & \checkmark & \checkmark & \checkmark & \checkmark\\
                                                &  \texttt{Homosexuality}  & \checkmark & \checkmark & \checkmark & \checkmark\\
                                                &  \texttt{Same-sex marriage}  & \checkmark & & \\
    \midrule
    \multirow{3}{*}{Religion}           & \texttt{Catholic Church}  & \checkmark & \checkmark & \checkmark & \checkmark\\ 
                                                &  \texttt{Buddhism}  & \checkmark & \checkmark & \checkmark & \checkmark \\
                                            & \texttt{Atheism}  & \checkmark  & \checkmark & \checkmark & \checkmark \\ 
    \midrule                                        
    \multirow{13}{*}{Technology}      &  \texttt{iOS}  & \checkmark & \checkmark & \checkmark & \checkmark\\
                                     &  \texttt{Android}  & \checkmark & \checkmark & \checkmark & \checkmark\\
                                    &  \texttt{Mac}  & \checkmark & \checkmark & \checkmark & \checkmark\\
                                    &  \texttt{Windows} & \checkmark  & \checkmark & \checkmark & \checkmark\\
                                    &  \texttt{iPhone X, 8, 8 Plus}  & \checkmark & \checkmark & \checkmark \\
                                   &  \texttt{Galaxy S8, S9}  & \checkmark & & \checkmark\\
                                         &  \texttt{Samsung Android}  & \checkmark & \checkmark  & \checkmark & \checkmark\\
                                         &  \texttt{Huawei} & & \checkmark & \checkmark & \checkmark\\
                                             &  \texttt{Oppo} & & & & \checkmark\\
                                         &  \texttt{Older devices}  & \checkmark & \checkmark & \checkmark & \checkmark\\
                                         &  \texttt{Smartphone and tablet}  & \checkmark & \checkmark  & \checkmark & \checkmark \\
                                 &  \texttt{Tablet}  & \checkmark & \checkmark & \checkmark & \checkmark \\
                                             &  \texttt{Technology early adopters} & \checkmark & \checkmark & \checkmark & \checkmark\\
    \midrule                                        
    \multirow{2}{*}{Connectivity}      &  \texttt{3G, 4G}  & \checkmark  & \checkmark & \checkmark & \checkmark\\
                                                &  \texttt{WiFi}  & \checkmark & \checkmark & \checkmark & \checkmark\\

    \bottomrule

\end{tabular}
\end{center}
\end{table*}

We build on previous studies to choose attributes of Facebook users that are predictive of socioeconomic status \cite{fatehkia2018using, fatehkia2020relative}. 
In particular, recent studies have shown that technology features such as the type and model of owned mobile devices, or the cell network used to access the Internet, are highly predictive of income \cite{fatehkia2020relative}.
We extend our study to a wider range of features, querying audience estimates over a range of 47 attributes that pertain to demographics, culture, mobility and other behaviors.
A complete list of attributes is reported in Tab. \ref{tab:features}.

How Facebook determines the audience that corresponds to a specific target is not disclosed in detail by the marketing platform. 
Some attributes are inferred by Facebook from the self-disclosed information and from the user interactions on the platform. 
Technology related features are automatically determined by the information collected from the devices used to connect to the platform, which in principle may be more reliable.

We queried the marketing API requesting the number of Facebook users matching the above targeting characteristics, and selecting two different age groups: users aged more than 13, which means all users of the platform, and only users aged more than 25.
As described below, we trained two distinct predictive models using the features of the two age classes and compared their predictive performance in all cities. 

%\section{Methods}

\subsection{Spatial aggregation of target and predictive features}
\label{sec:spatial}

%%% FIGURE %%%%
\begin{figure}[t]

  \centering
  \includegraphics[width=\columnwidth]{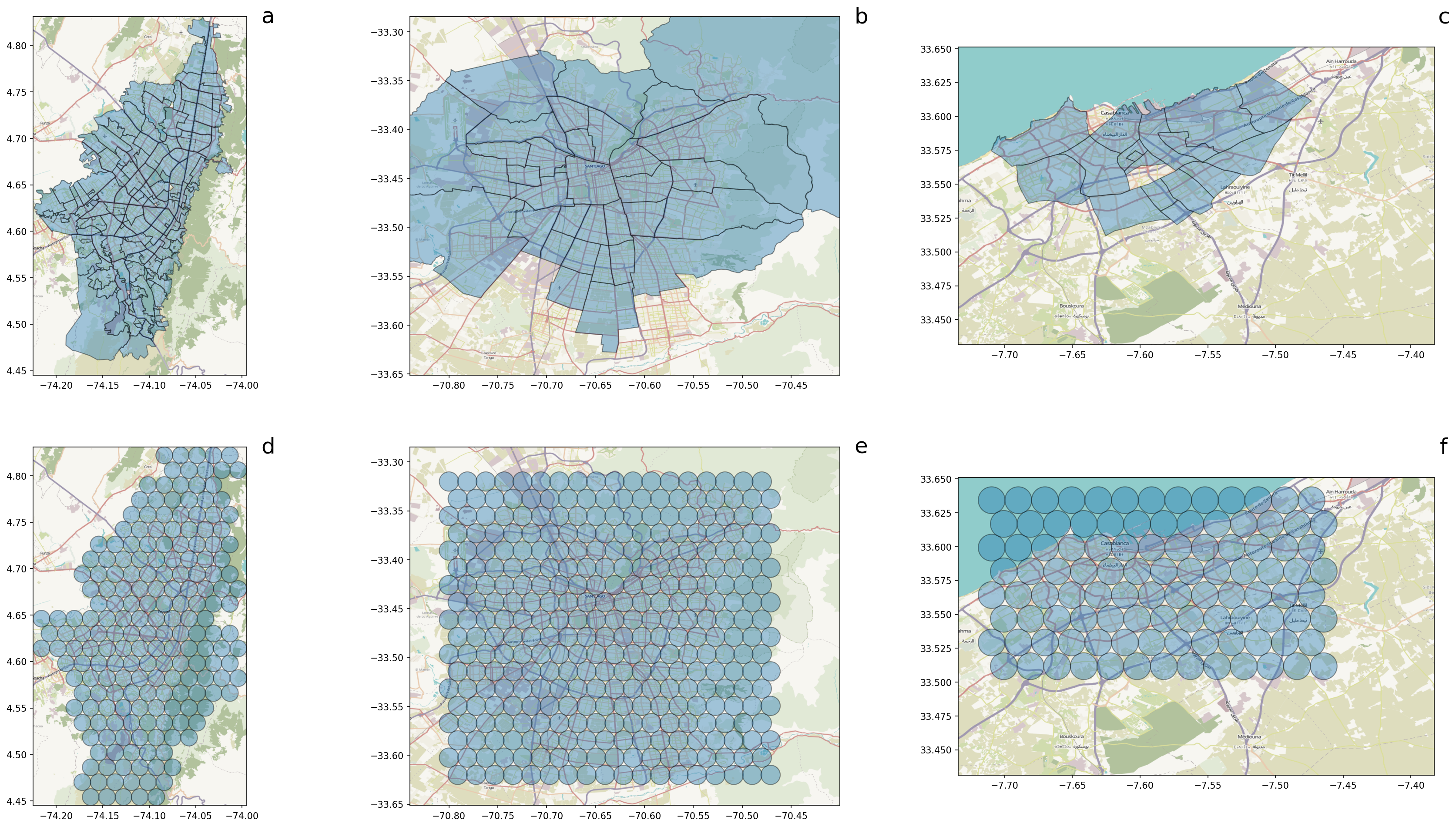}
  \caption{\label{fig:spatial} Spatial targeting schemes for the queries of Bogota (a, d), Santiago (b, e), Casablanca (c, f). For each city we show the administrative subdivisions for which socioeconomic data is available (upper panels) and the grid of circles of 1 km radius that we used to collect audience estimates. Only estimates for circles intersecting the administrative units were considered in the models.}
\end{figure}

As mentioned before, the Facebook marketing API provides audience estimates for a specific geographic area that is specified in each query. Available geographic targeting options vary in each country and our queries used different spatial parameters, depending on the city under study.

In the city of Atlanta, and in general for all US cities, the Facebook marketing API allows to run queries by zip code, that is the spatial granularity for which income data is available. 
In this case, it was possible to match the predictive features of the audience estimates with the target variable, at the same spatial resolution, without the need of additional data manipulation.
In other countries, however, the platform provides fewer geographic targeting options, especially at a high granularity.
To overcome this issue, in the case of Bogotá, Santiago, and Casablanca, we run our queries by selecting circles of 1 km radius as our geographic target.
This corresponds to the highest resolution at which Facebook audience estimates are available from the marketing API.

We created a grid of equally spaced circles of 1 km radius to cover the area of the three cities as shown in Fig.~\ref{fig:spatial}. 
We then queried the Facebook API by requesting the number of users who live in every circle and who match the targeting options defined above. 
Finally, we projected the audience estimates from the circles to the administrative units as follows.
For each circle $i$, and each administrative unit $j$ (\emph{comuna}, UPZ or \emph{arrondissement}) we compute the fraction of the area of the circle $i$ that intersects the unit $j$, $a_{ij}$. 
Then, we compute the estimate of MAU in each unit $j$ as the sum of the estimates in the circles intersecting $j$, weighted by the area of intersection: 
\begin{equation*}
MAU_{j} = \sum_{i \in \nu_{j}} MAU_i \; a_{ij}    
\end{equation*}
where $\nu_{j}$ is the set of circles intersecting the administrative unit $j$.

\subsection{Predictive model}

For each city we considered Facebook attributes counts normalized by the total number of Facebook MAUs, in each administrative unit, as input predictors for a regression model. 
We defined two different regression models, using Facebook attributes of two age classes: the first model considers the attributes of all Facebook users (aged 13 or more) and the second model considers only users 25 and older. 
Tab. \ref{tab:features} reports with a check mark whether a Facebook attribute was included or not in the input matrix of the model without age limits, for each city, before variable selection.

In both cases, the design matrix $\boldsymbol{X} \in \mathbb{R}^{n\times p}$ is computed as follows: (i) each row corresponds to a neighborhood (at the chosen aggregation level) with all normalized Facebook variables greater than zero, and (ii) each column corresponds to a feature without zero entries in every district of the city. 
With the pair $(\boldsymbol{X}, \boldsymbol{y})$, where $\boldsymbol{y} \in \mathbb{R}^n$ is the target vector of  socioeconomic indicator, we train a Lasso estimator to predict $\boldsymbol{y}$ from $\boldsymbol{X}$ and to select explanatory features. 
The model is tuned by searching for the best hyperparameter $\alpha$ from a set of candidates spaced on a log scale, with a 5-fold cross validation optimizing the Lasso objective with respect to the variables $\boldsymbol{\beta} \in \mathbb{R}^{p}$ in each fold:
\begin{equation*}
\hat{\boldsymbol{\beta}} = \mathrm{argmin}_{\boldsymbol{\beta} \in \mathbb{R}^{p}} \Big\{\frac{1}{n} || \boldsymbol{y} - \boldsymbol{X}\boldsymbol{\beta} ||_2^2 + \alpha ||\boldsymbol{\beta}||_1 \Big\}
\end{equation*}
The goodness of fit is measured with $R^2$ score evaluated between the real socioeconomic indicators and cross-validated estimates. 
The chosen $\alpha$ value is the one that has the best average $R^2$ over 30 different cross validation splits. 
The variables selected by the best model are then used as input to standard linear regression.

% %%% %%% %%% %%% %%% %%% %%% %%% %%% %%% %%% %%% %%% %%% %%% %%% %%% %%% %%%
%%% %%% %%% %%% %%% %%% %%% %%% %%% %%% %%% %%% %%% %%% %%% %%% %%% %%% %%% %
% %%% %%% %%% %%% %%% %%% %%% %%% %%% %%% %%% %%% %%% %%% %%% %%% %%% %%% %%%

\section{Results and discussion}

\begin{table*}
\begin{center}
  \caption{Performance of linear regression models fitted on selected variables. The $F$ indicates the number of features that are used as input to each model, before variable selection. Reported values are $R^2$ based on a leave-one-out cross validation.}
  \label{tab:scores}
  \begin{tabular}{lcccc}
    \toprule
   City & \multicolumn{2}{c}{$F$} & \multicolumn{2}{c}{$R^2$}\\
     & Model All Ages & Model $>$25 & Model All Ages & Model $>$25\\
    \midrule
    Atlanta, GA, USA & 44 & 43 & 0.4252 & 0.5596\\
    Bogotá, Colombia  & 41 & 40 & 0.5117 & 0.5696\\
    Santiago, Chile  & 44 & 41 & 0.8277 & 0.9323\\
    Casablanca, Morocco & 38 & 37 & 0.0231 & 0.4646\\
  \bottomrule
\end{tabular}
\end{center}

\end{table*}

The predictive performance of the regression models is shown in Tab. \ref{tab:scores}, where cross-validated $R^2$ values are reported for each city, along with the number of features that are used as input before variable selection.
Although the number of Facebook attributes available for the regression differs across cities, model's performance does not significantly depend on the number of available features.

In all cases, performance improves when the model is trained on the Facebook estimates of adults users only, those aged 25 or more. 
In the case of Casablanca, the improvement is especially drastic, leading to $R^2 = 0.46$ from an initial $R^2 = 0.02$.  
In the other cities, the improvement is more limited but, overall, considering only users above 25 leads to an average 18\% increase in $R^2$.
The fact that restricting queries to adult users provides a better predictive performance of socioeconomic status highlights the relevance of users' age profile when dealing with non-traditional data sources. 
Interestingly, when considering only adult users, the model achieves similar results in both developed and developing countries, as demonstrated by the case of Atlanta and Bogotá, where the model equally achieves $R^2 = 0.56$. 
For the city of Santiago, the performance of the model is exceptionally good, achieving $R^2 = 0.93$ with a model based on adult users' features only.
The difference in performance may be due to the Facebook penetration in each region: Morocco at 55\%, Colombia 63\%, USA 70\%, and Chile 79\% (Facebook users as a fraction of the total population).
Thus, higher Facebook penetration leads to an overall better model performance, still, the range of the performances is wider than that of Facebook penetration, pointing to the sensitivity of the included variables.

\begin{figure}[t]

  \centering
  \includegraphics[width=\columnwidth]{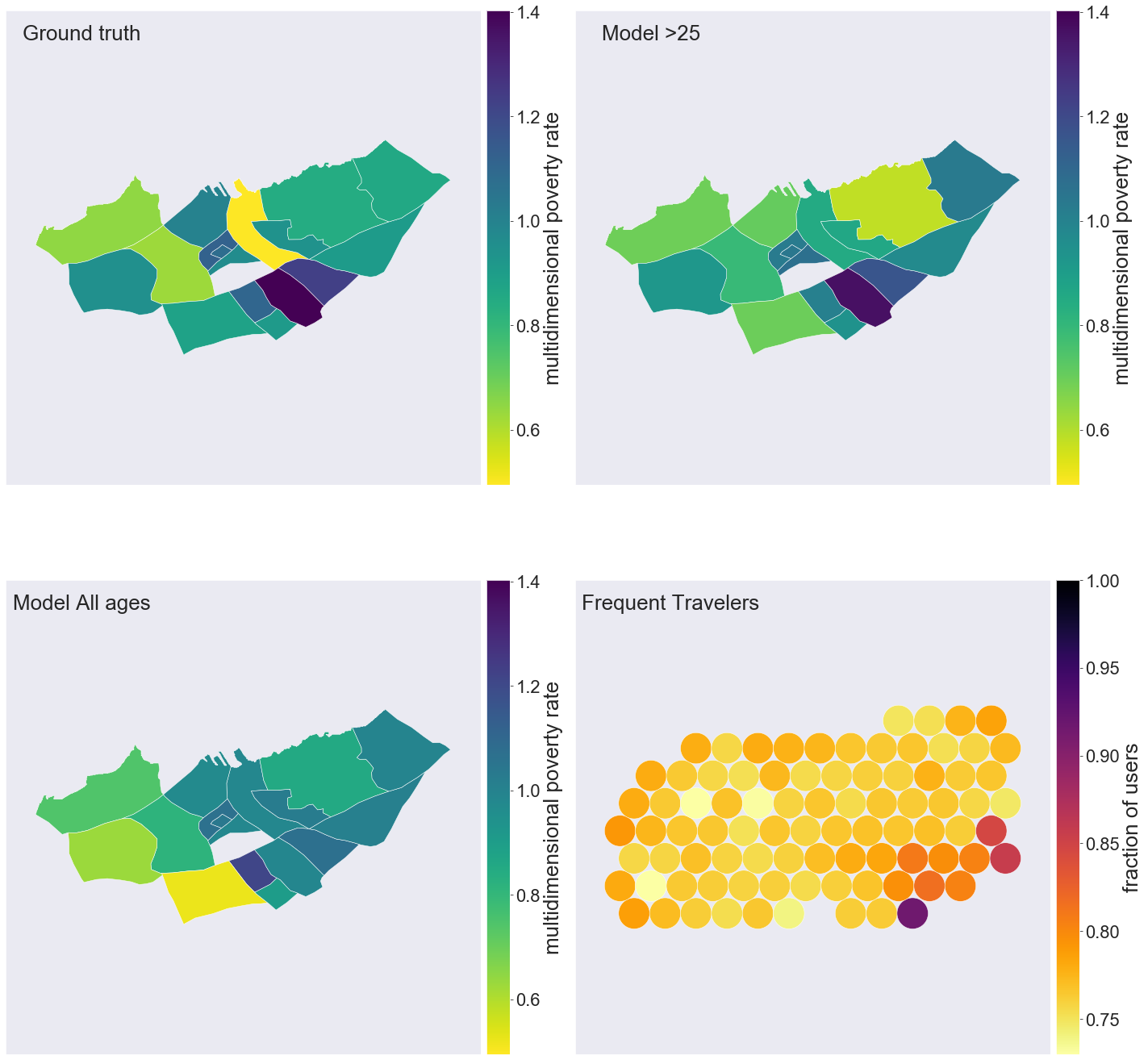}
  \caption{\label{fig:casablanca} Poverty estimates for Casablanca: upper-left: ground truth by district, upper-right: prediction using Facebook ads estimates for adults aged more than 25, lower-left: model prediction using users of all ages, lower-right: proportion of frequent travelers in Facebook users (age $>$25).}
  
\end{figure}

Recall that, beside Atlanta, the variables were obtained on the level of a 1~km radius circle grid, potentially providing a finer-grained view of the area.
For instance, in Fig. \ref{fig:casablanca} we show four plots for Casablanca: the ground truth, the output of our models using all ages, or only $>$25, and the circles' grid of values for the Frequent Travelers variable. 
First, we observe the marked improvement of the model when only the adults are considered.
Second, we find a variability among the areas on the scale of the circles, outlining a neighborhood on the lower right with dark circles signifying lower socioeconomic status, and highlighting pockets of higher one in center-left. 
Such fine-grained resolution allows to go beyond politically defined areas, and potentially reveal variability that is hidden by aggregation over large areas.

To gain further insights into the Facebook attributes that are predictive of socioeconomic status of neighbourhoods of the four cities, in Tab. \ref{tab:score_variable} we show the variables selected by each model and the associated L1-norm regularized coefficients $\beta_i / ||\boldsymbol{\beta}||_1$.
Positive values of the coefficients are associated with a higher level of socioeconomic status, with the exception of Casablanca where the target variable is the multidimensional poverty rate and, therefore, positive values of $\beta_i$ are associated with higher poverty rates.

\begin{table*}
  \caption{Model selected variables for the prediction of socioeconomic status. Each row indicates the Facebook variables that are selected by the model to predict the socioeconomic indicators in each city. Variables selected by the All age model and the >25 model are shown separately.} 
  \label{tab:score_variable}
  \begin{tabular}{lllr@{~}l}
  %\begin{tabular}{lllrl}
    \toprule
    City & Model & Selected Variable & \multicolumn{1}{c}{$\frac{\beta_i}{||\boldsymbol{\beta}||_1}$} \\
    \midrule
    %%%ATLANTA
    \multirow{17}{*}{\rotatebox[origin=c]{90}{Atlanta, GA, USA}}          &\multirow{7}{*}{All Ages}& \texttt{Married} &\PosbarErr[0.081]{0.025}\\
    & & \texttt{WiFi}&\PosbarErr[0.048]{0.026}\\
    & & \texttt{iOS}&\PosbarErr[0.029]{0.024}\\

    & & \texttt{High school grad}&\NegbarErr[-0.067]{0.0478}\\
    & & \texttt{Android}&\NegbarErr[-0.071]{0.024}\\
    & & \texttt{Casino} &\NegbarErr[-0.170]{0.074}\\
    & & \texttt{Engaged} &\NegbarErr[-0.485]{0.235}\\
    
    \cline{2-4}
    &\multirow{10}{*}{$>$25}& \texttt{Females}&\PosbarErr[0.093]{0.070}\\
    & & \texttt{Engaged}&\PosbarErr[0.079]{0.038}\\
    & & \texttt{iOS}&\PosbarErr[0.026]{0.018}\\

    & & \texttt{Cooking} &\NegbarErr[-0.048]{0.027}\\
    & & \texttt{Same-sex marriage} &\NegbarErr[-0.069]{0.051}\\
    & & \texttt{Older devices}&\NegbarErr[-0.074]{0.055}\\
    & & \texttt{Mac} &\NegbarErr[-0.077]{0.034}\\
    & & \texttt{Galaxy S8} &\NegbarErr[-0.083]{0.053}\\
    & & \texttt{Smartphone e tablet} &\NegbarErr[-0.136]{0.101}\\
    & & \texttt{Windows} &\NegbarErr[-0.164]{0.121}\\
    
    \midrule
    
    %%%BOGOTA
    \multirow{10}{*}{\rotatebox[origin=c]{90}{Bogotá, Colombia}}           &\multirow{5}{*}{All Ages}& \texttt{Buddhism} & \PosbarErr[0.353]{0.141}\\
    & & \texttt{Frequent international travelers}&\PosbarErr[0.089]{0.064}\\
    & & \texttt{Away from hometown}&\PosbarErr[0.052]{0.141}\\
    & & \texttt{Huawei} &\NegbarErr[-0.072]{0.026}\\
    & & \texttt{Engaged} &\NegbarErr[-0.407]{0.110}\\
    \cline{2-4}
    
    &\multirow{5}{*}{$>$25}& \texttt{Frequent international travelers}&\PosbarErr[0.344]{0.083}\\
    & & \texttt{iOS}&\PosbarErr[0.141]{0.090}\\
    & & \texttt{High school grad} &\PosbarErr[0.083]{0.247}\\
    & & \texttt{Huawei} &\NegbarErr[-0.099]{0.045}\\
    & & \texttt{Engaged} &\NegbarErr[-0.285]{0.116}\\
    
    \midrule
    %%%SANTIAGO
    \multirow{13}{*}{\rotatebox[origin=c]{90}{Santiago, Chile}}           &\multirow{9}{*}{All Ages}& \texttt{iPhone 8} & \PosbarErr[0.559]{0.160}\\
    & & \texttt{Married}&\PosbarErr[0.093]{0.027}\\
    & & \texttt{College grad}&\PosbarErr[0.047]{0.027}\\
    & & \texttt{Older devices} &\NegbarErr[-0.045]{0.048}\\
    & & \texttt{Males} &\NegbarErr[-0.060]{0.044}\\
    & & \texttt{Engaged} &\NegbarErr[-0.062]{0.176}\\
    & & \texttt{Technology early adopters} &\NegbarErr[-0.072]{0.187}\\
    \cline{2-4}
    &\multirow{7}{*}{$>$25}& \texttt{iPhone 8} &\PosbarErr[0.627]{0.120}\\
    & & \texttt{Married}&\PosbarErr[0.072]{0.019}\\
    & & \texttt{College grad} &\PosbarErr[0.046]{0.021}\\
    & & \texttt{iOS} &\PosbarErr[0.036]{0.021}\\
    % & & \texttt{tablet} &\NegbarErr[-0.002]{0.018}\\
    & & \texttt{Males} &\NegbarErr[-0.068]{0.027}\\
    & & \texttt{Casino} &\NegbarErr[-0.119]{0.045}\\
    \midrule
    
    %%%CASABLANCA
    % \multirow{8}{*}{Casablanca, Morocco}           &\multirow{5}{*}{All Ages}& \texttt{Technology early adopters} & \PosbarErr[0.362]{1.553}\\
    \multirow{8}{*}{\rotatebox[origin=c]{90}{Casablanca, Morocco}} &\multirow{5}{*}{All Ages}& \texttt{Technology early adopters} & \PosbarErr[0.362]{1.1}\\
    & & \texttt{Windows}&\PosbarErr[0.197]{0.475}\\
    & & \texttt{High school grad}&\PosbarErr[0.175]{0.520}\\
    & & \texttt{Engaged}&\PosbarErr[0.156]{0.404}\\
    & & \texttt{Females}&\NegbarErr[-0.094]{0.106}\\
    \cline{2-4}
    &\multirow{3}{*}{$>$25}& \texttt{Returned from travels 1 week ago} &\PosbarErr[0.796]{0.246}\\
    & & \texttt{Frequent Travelers}&\PosbarErr[0.037]{0.063}\\
    & & \texttt{Android} &\PosbarErr[0.026]{0.016}\\

    \bottomrule

\end{tabular}
\end{table*}

In line with previous studies, technology related features are often selected as predictive of SES. 
Corroborating findings in previous works \cite{rama2020facebook,fatehkia2020relative}, Apple products (iOS and iPhone) are reliably associated with higher SES.
On the other side, use of Android devices is consistently associated with lower income.  
However, other variables are also often selected by the models: estimates of educational attainment are found in several models, as well as demographic attributes including gender and marital status. 
Though they may have different association with the SES, for instance being engaged has a positive coefficient for Atlanta, but negative for Bogotá. 
Notably, a higher prevalence of female Facebook users is associated with higher socioeconomic development in both Atlanta and Casablanca, while a higher proportion of male users is predictive of lower income in Santiago. 
These results confirm in a urban setting the association between the Facebook gender gap and development indicators, which has been previously observed at country level \cite{fatehkia2018using}. 
Overall, these findings suggest that SES modeling can be improved by considering a broader selection of variables than just the technological ones (as used in \cite{fatehkia2020mapping}), improving both the model fit and expanding the set of correlates with SES.

\subsection{Advantages}

Above, we illustrate the feasibility of using Facebook advertising platform as a source of fine-grained socioeconomic information in urban areas around the world. 
This data source has several advantages over the traditional survey methods. 
First, it is publicly available and it is possible to gather large amounts of data via the website's API. 
Second, it is updated regularly, and may provide a more up-to-date view of the situation than an expensive census or survey. 
Third, disaggregation by gender and age provide a way to focus on target demographics of interest, such as in previous work on gender inequality in India \cite{mejova2018measuring}.
Fourth, because the individual data is not released by Facebook, this resource allows the study of populations without the compromise of privacy of any captured individuals. 
Fifth, this data source may reveal populations which are not officially counted by the local authorities, or who are temporarily passing through the area, such as recent study of the Venezuelan migration into Colombia \cite{palotti2020monitoring}. 
Finally, it is possible to explore the demographic, behavioral, and technological correlates of socioeconomic index in each urban setting. For instance, in previous study of the Italian municipalities, those with lower income had a higher interest in cooking, restaurants, and gambling \cite{rama2020facebook}. Although not explored in this work, health-related interests may also help identify areas of need \cite{araujo2017using}.
In our case study, we provide an analysis of four cities of varying wealth dynamics, Facebook penetration, and part of the world. 
We illustrate that the signals provided by Facebook advertising platform are indeed related to socioeconomic indicators, and in fact may provide a finer-grained detail on the separation of their inhabitants by wealth.

%Satellite imagery
As a comparison, several alternative data sources have been proved to be effective to measure economic development. 
In particular, thanks to the recent advances in image processing and machine learning, information extracted from satellite areal imagery represents one of the most widely investigated resource to map SES at different scales \cite{jean2016combining, burke2021using}. 
Satellite-based measurements can achieve a very high predictive accuracy combined with a high spatial resolution. 
However, such levels of accuracy come at a significant financial cost since high-resolution (< 1$m$) satellite imagery must be purchased from private providers and it is usually expensive.  
Also, satellite-based measurements of development often lack interpretability and such issue has been addressed only recently \cite{abitbol2020interpretable, ayush2020generating}.   
Compared to satellite images, social media advertising data are usually less expensive to collect, and their relationships with SES are easier to interpret. 
Also, social media data may be more suitable to capture socio-demographic changes that may reflect changes of SES on short timescales. 
Combining the two data sources, satellite imagery and advertising data, may provide complementary information to advance SES mapping at a high granularity.  

\subsection{Limitations and Future work}

The above advantages come with marked limitations, which must be addressed when utilizing this data source.
The dynamic nature of this data reminds us that Facebook may update it based on the internal scheduling and needs of the company, and it is not certain just how current the estimates are. 
Further, the black-box nature of the tool puts in question whether identification of individuals in various categories performs uniformly across locales. For instance, whether the gender classification (when such information is not provided by the user) works equally well for African, Asian, and Middle-Eastern users as it does for English-language ones is questionable, given the known biases of ``minority language'' NLP systems \cite{blodgett2020language}. 
However, if we study fairly homogeneous populations within each locale, the analytical pipeline applied to the users will hopefully not be as subject to such bias as a comparative study across countries or language groups.
Another source of bias may come from the internal benefit to the company to find particular users that are highly sought-after by the advertisers, such as those having the funds to spend on the advertised product, or having the demographics matching the advertised messages.
Further, despite the privacy-preserving nature of the tool, there may be situations in which potentially vulnerable populations may be identified. For instance, in the above-mentioned study on the Venezuelan migration, the governments of both Venezuela and Colombia may use this information to reinforce certain border crossings or deploy more police presence in certain areas \cite{palotti2020monitoring}. 
We urge the research community to apply the same ethical standards of research to this new data source as would be applied to any standard research methodology.

Beyond the limitations of the data source, this study in particular has notable shortcomings, some of which present interesting future research directions.
We present only one way of using the 1~km radius circles to survey an area, but other packing and aggregation methods may be possible. For example, when projecting from circles to area, the geographic overlap may be enriched by the population data of the two areas.
Further, it is difficult to make generalizations based on 4 cities, and we are planning on performing similar experiments in more cities, including in low-income economies (unfortunately the availability of the fine-grained SES ground truth data is often the limiting factor). 
As described above, we will utilize other data sources, such as satellite imagery, as a potential alternative for comparison. Note that, a previous work on using satellite imagery for SES modeling in Santiago shows lower $R^2$ scores (0.691) than achieved by our model \cite{piaggesi2019predicting}. 
Finally, we are looking forward to strengthening the theoretical underpinning for the selection of model features, both to enrich our understanding of the correlates of SES, and to model other wellbeing factors (such as those associated with physical and mental health, access to services, etc.).

\section{Conclusion}

In this study, we modeled the socioeconomic indicators across four cities situated in different continents and undergoing different economic development stage. 
We showed that, using Facebook advertising estimates, it is possible to obtain fine-grained models of SES of populations in the urban areas of Atlanta (GA, USA), Santiago (Chile), Bogotá (Colombia), and Casablanca (Morocco). 
For each city, we show that a different set of demographic, technological and behavioral variables may be associated with SES, and that it is helpful to model working-age adults when collecting Facebook data. 

Using methodology proposed here, we hope that the SDG goal of poverty reduction will be monitored at a fine spatial resolution in the urban areas worldwide, both to gauge the improvements in socioeconomic factors, and to better understand the multiple dimensions of wellbeing. 
As we continue to build such systems, we encourage researchers and policy-makers to continue experimentation with this near-real-time, fine-grained data source, especially in the dynamic urban environments of the developing world.

\section{Acknowledgments}
SG, SP, YM, AP and MT gratefully acknowledge the support of the Lagrange Program of the ISI Foundation funded by CRT Foundation. 
MK acknowledges to participate as the Fellow of the ISI Foundation and support from the H2020 SoBigData++ project (H2020-871042) and the DataRedux ANR project (ANR-19-CE46-0008).
We acknowledge Ingmar Weber for sharing code to access the Facebook marketing API.
We are thankful to Yanyan Xu, Luis Eduardo Olmos and Marta Gonzalez for help with accessing the Colombian socioeconomic data.

%\bibliographystyle{ACM-Reference-Format}

\bibliography{references}

\end{document}